\documentclass[preprint2]{aastex631}

\begin{document}

\title{The Celestial Vault is Not a Dome: Implications for the Moon Illusion}

\author{Karl Hipius}
\author{David A. Kornreich}
\affiliation{%
Department of Physics, State University of New York at Cortland, Cortland NY 13045
}%

\begin{abstract}
The moon illusion, in which the moon appears larger at the horizon than at higher altitudes, has been investigated since antiquity, yet it remains not fully explained.
Our method of investigating the phenomenon is based on that attributed to Martin Folkes \citep{SM}. Folkes suggests investigators identify landmarks on the ground and request human subjects indicate a place in the sky that they perceive to be above that location by raising their arm and pointing. The intersection of a vertical line rising out of the landmark and the ray extending from the subject's finger then identifies a point on the subject's model of the celestial vault. When repeated for landmarks covering a range of distances, the subject's entire celestial vault can be traced. We asked 30 subjects to perform such an identification on a series of points between a horizontal distance of 3~and 12600~meters across featureless open water in Verplanck and Ithaca, New York. The resulting data are in disagreement with the widely presumed flattened dome model, and, in particular, the Size--Distance Invariance Hypothesis. 
In addition, we find that the vault does not intersect the ground and leaves an indeterminate space of about $20^\circ$ altitude above the horizon with no correspondence to the ground. It is in this region that the Moon Illusion is most pronounced.
\end{abstract}

\section{\label{sec:intro}Introduction}
To the casual observer, the Moon often appears larger at the horizon and smaller at altitude. This effect occurs despite the fact that the physical size of the moon is of course identical in the two configurations and its angular sizes are equal to within the naked eye's ability to perceive them. This illusion has been known since antiquity, and yet remains not fully explained. Even the definition of ``appearing larger'' is still at issue: is the illusion a mistake of angular size, of linear size, or of distance? 

Three classes of theories explaining the illusion have attempted to explain the perception. The first of these is physiological, involving the position and movement of the observer's head and eyes. \citet{HB} found when raising the eyes to look at the moon, its apparent size would shrink, but the effect disappeared when observers were asked to keep their eyes still and tilt their head back instead. However, \citet{KAR} found ocular position not to be a significant contributor to the illusion. Similarly, \citet{BGS62} found insignificant effects of head and eye position. 

Ultimately, there must be some effect of head position as the illusion disappears when viewed upside down \citep{HAD}. Moreover, more modern studies have implicated binocular vergence with the illusion \citep{Kim,Suzu}, and the illusion is known to be minimized when viewed with only one eye \citep{TB}.

The second of the three classes of theories is that of relative size, in which the moon's size is said to be determined by 
context clues such as trees and buildings in the same field of vision. This theory hypothesizes that the apparent size of the moon depends on the visual size of the objects surrounding the moon. In other words, when the Moon is ``surrounded'' by objects in the nearby visual field whose size 
an observer is familiar with, it appears comparably very large. However, when the moon is at a higher altitude in the sky, only the clouds and the night sky can be used for size comparison and therefore it appears much smaller. 

In the literature, 
there has been disagreement as to whether or not the terrain changes the apparent size of the moon. \citet{KAR} found terrain is essential for the illusion to occur whereas \citet{Suzu} obtained a different effect, whereby the illusion still appeared in complete darkness with no visible terrain. It should be noted that Suzuki's successful measurement in the complete darkness of a planetarium leaves only physiological considerations as important. 

Finally, apparent distance theories were first suggested by Ibn al-Haytham in the 11th century, and more recently tested by \citet{RK1}. These theories most commonly suggest apparent size of an object is dependent upon apparent distance. They measured apparent size difference between the zenith and horizon moon and they found a wide range of differences between individuals, but ultimately found that on average, the horizon moon appears approximately 34 percent larger than the zenith. In this model the celestial vault is perceived as a two dimensional plane or a flattened dome. If this is true, then for the moon to appear larger on the horizon, it must also appear as further away from the observer. This is a corollary of Emmert's Law \citep{KG} whereby the perception of an object's size is proportional to the perceived distance to the object. This fails to explain, however, why observers report the moon to actually appear closer when at the horizon as compared to the zenith. This has been tested by \citet{BW}. If the flattened celestial vault hypothesis were accurate, it would also be expected that the moon would appear larger when near the horizon and not as large at the zenith. Their results did not support the flattened celestial vault hypothesis. 

Thought to be intimately related to the Moon Illusion is the perceptual effect known as \emph{size constancy,} in which the adjusted size of an object remains constant despite changes in the size of its physical image on the retina. Size judgments are made, therefore, using some combination of measured angular size and perceived distance. Thus the most well--known theories of the Moon Illusion are those which incorporate the size--distance~invariance hypothesis (SDIH) that the perceived metric size and perceived distance of an object are directly related. In other words, for a body with given constant angular size such as the Moon, a larger perceived distance would result in a larger perceived metric size. 

\begin{figure}[tp]
    \centering
     \includegraphics[width=\linewidth]{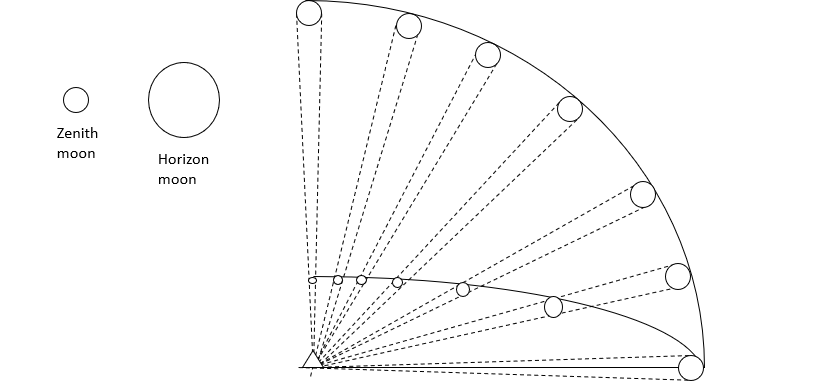}\\
    \caption{Smith's model of the perceived shape of the sky. As the Moon's altitude increases, its perceived distance decreases because of the flattened dome, and with decreasing distance, decreasing judgements of metric size. \citet{SM}\label{SmithFigure}}
\end{figure}

The SDIH, combined with the perceived enlargement of the Moon, implies that the observer judges the Moon to be more distant at the horizon, despite the true, physical distance actually being about 2\% nearer. 
The observer's mind must therefore be applying a mental model of the shape of the sky during image processing. In this paper we will use the phrase ``celestial vault'' to refer to a human subject's mental model of the shape of the sky. In the 11th century, Ibn al-Haytham is the first known philosopher to have attempted to describe this model by describing the celestial vault as a plane above the observer. 
Later, \citet{Des} developed the first theory involving a mental model in the shape of a flattened dome. 
Figure~\ref{SmithFigure} illustrates the general idea \citep{SM}. 
Perceived metric size and perceived distance to an object are directly proportional. For the moon to appear larger on the horizon, it must appear more distant at the horizon than near zenith, implying that the celestial vault is a flattened dome. In each sector of the figure, the true constant distance to the moon is represented by the outer moon whereas the apparent distance to the moon is represented by the inner moon which travels along a more flattened sphere. The real outer moons' constant distance creates a constant image size on the observer's retina; however, the perceived inner moons' varying distance creates perceptions of varying sizes at constant angular diameter.



\begin{figure}[tp]
    \centering
     \includegraphics[width=\linewidth]{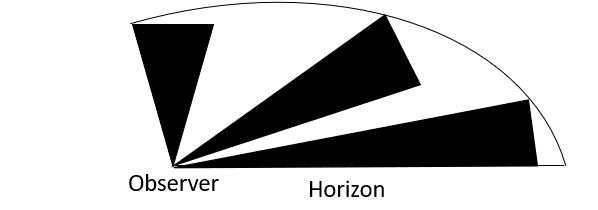}\\
    \caption{Emmert's Law, demonstrating on a flattened dome how equal angles can correspond to psychologically larger or smaller bodies. Bodies located on the lower altitude, more psychologically distant part of the sky are perceived as larger. \citet{KG}\label{EmmertsLawFigure}}
\end{figure}

\citet{KG} used participants' retinal after--images as substitute moons as a way to exclude retinal image size as a source of the illusion. They determined that the illusion still occurs when using such after--images, i.e., an after--image on the horizon appears larger than itself at zenith. Emmert's Law, that increasing the judged distance to an object subtending a constant visual angle increases its perceived size, has the ability to predict this effect. The implication is that the sky is indeed perceived as being more distant at lower altitudes. This is depicted in Figure~\ref{EmmertsLawFigure}.

There have been relatively few attempts to actually measure the degree of flattening of the perceived sky. Smith, and more recently \citet{MN}, sought to determine the degree of flattening by measuring the perceived bisector of the horizon--zenith angle. 
In other words, this bisector should be a line that divides the horizon--zenith angle into two equal parts.
A hemispherically shaped sky should obviously yield a value of 45$^\circ$, but when participants are asked, they consistently underestimate this angle by as much as 15$^\circ$. Other estimates have yielded underestimates of $25^\circ$ \citep{ANG}, consistent with a very flattened sky.  

 \citet{Rees} generalized the method of sky bisectors and asked participants to determine apparent elevation angle of the Sun and Moon at various altitudes. Rees utilized locations with a distant, flat, unobstructed view of the horizon on a clear day, and determined that altitude was consistently overestimated, especially in the range of $35^{\circ}$ to $45^{\circ}$, in agreement with the flattened celestial vault and SDIH. This differs from the method used by Smith as well as \citet{MN} because it is measuring the angle to objects, the Sun and Moon, instead of the night sky.

\citet{BW} conducted three experiments, one asking participants to determine relative distances to various points in the sky, a second asking participants to estimate the relative distance between the zenith and horizon, and the final experiment essentially repeating the second experiment but replacing the horizon with the roof line of nearby buildings of known physical distance. They ultimately found no evidence of either a flattened dome or an anomalously distant horizon.

More recently, \citet{T1} investigated the role of direction on distance estimates to objects. Participants in a dark room were shown three dim rectangular lights in various viewing directions between horizontal and zenith, and asked to move them until their distances from the observer were perceived to be equal. For distances greater than 1~m, shorter vertical distances were perceived to be equal to longer horizontal distances. This was interpreted to indicate that for images at the zenith, the celestial vault is actually elongated vertically, in contrast to the flattened dome hypothesis.  

\begin{figure}
    \centering
     \includegraphics[width=\linewidth]{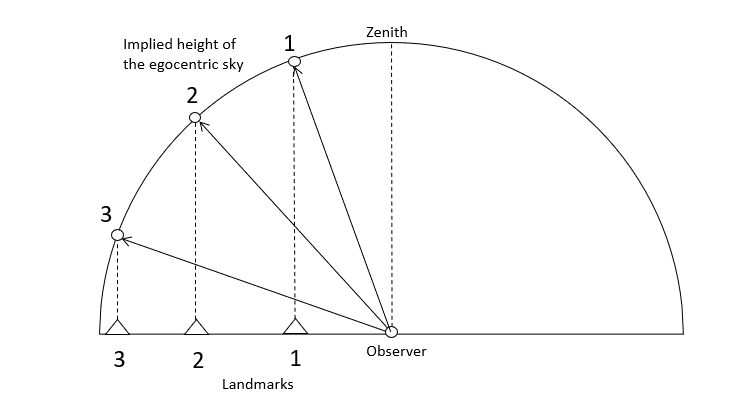}\\
    \caption{Using landmarks to map the perceived shape of the sky. Observers point towards points they perceive to be directly above landmarks. The intersection of their indication and a vertical from the landmark defines their celestial dome.\label{Book}}  
    
\end{figure}

The purpose of the experiment of this paper is to determine the shape of the celestial vault directly. The method we will use is the method Smith originally attributes to Martin Folkes, in which we use a series of fixed points on the ground at various known distances from the observer, who indicates corresponding points in the sky judged to be directly above each of these (See Figure~\ref{Book}).

The intersection of the verticals from the landmarks and the angle of altitude indicated by the observer yields the observer's model of the sky.


 \section{\label{sec:Method}Methods}
Our method is based on that proposed by Martin Folkes \citep{SM}. Folkes suggests investigators identify locations on the terrain and request subjects point to a place in the sky that they perceive to be above that location. Data were collected at two localities, one at Stewart Park in Ithaca, NY and the other at Cortlandt Waterfront Park in Verplanck, NY. Both of these locations were selected based on their proximity to water. Subjects in Ithaca were asked to look north over Cayuga Lake; those in Verplanck to look west over the Hudson River. A line of sight over a large body of water guarantees observation over a level surface without interference from hills, uneven land, or intervening objects between the subject and the distant horizon. Observations were made on cloudless days or when no more than approximately one twentieth of the sky was covered by wispy cirrus clouds. On such days, observations were only taken when no clouds were visible along lines of sight being tested. All observations were made in full daylight, with the Sun higher than 20~degrees of altitude above the horizon. 

Observers were asked to identify eight predetermined landmarks at the Ithaca location and ten predetermined landmarks at the Verplanck location. The distance from each subject to each landmark was measured using either a tape measure for very nearby landmarks, or Google Earth for landmarks beyond 20~meters. Landmarks were specifically chosen to avoid line of sight being restricted to a single direction. Instead, subjects were asked to identify widely separated landmarks around the subjects' horizon, so that each successive observation would alter body and eye position in an effort to avoid cognitive interference from one observation to the next. The first observations were those made in Ithaca. However, the most distant landmark that could be seen over water at that location was only approximately 8000~meters away. 
This distance is shorter than 11~kilometers, the typical distance to the horizon. Because data were not conforming to expectations in \citet{SM} and subsequent works, we then sought out a location from which more distant landmarks could be found, and settled on the Verplanck location, where the farthest visible landmark was approximately 13000~meters away. 

For each observation, date, time, and the subject's age and gender were collected. Then subjects were read the following instructions: \emph{``When I point out a landmark with my extended arm, I would like you to look up to the sky in the direction of that landmark and point out to me with your extended arm, a point in the sky that you perceive to be directly above that landmark.''} When subjects pointed, the angle of their extended arm above the horizontal was measured with an inclinometer. Prior to their respective observations, subjects had not seen previous trials with other subjects.

 \section{\label{sec:Data}Results}
\begin{figure}
    \centering
     \includegraphics[width=\linewidth]{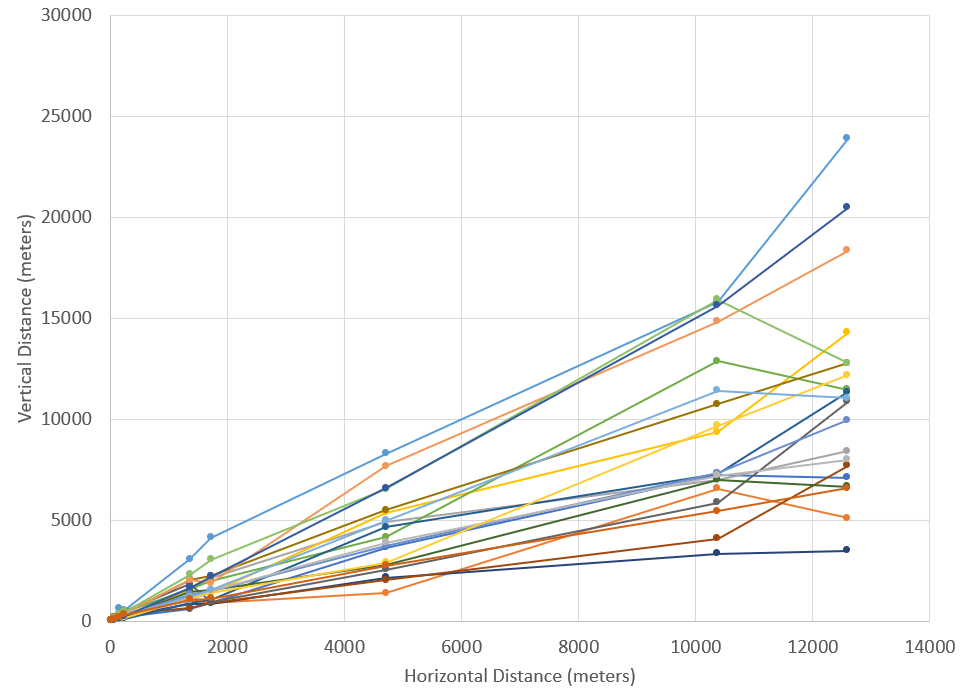}\\
    \caption{Verplanck Data: Note these are physical horizontal distances and derived vertical distances} \label{VerpFit}
    \vspace{2mm}
\end{figure}

\begin{figure}
    \centering
     \includegraphics[width=\linewidth]{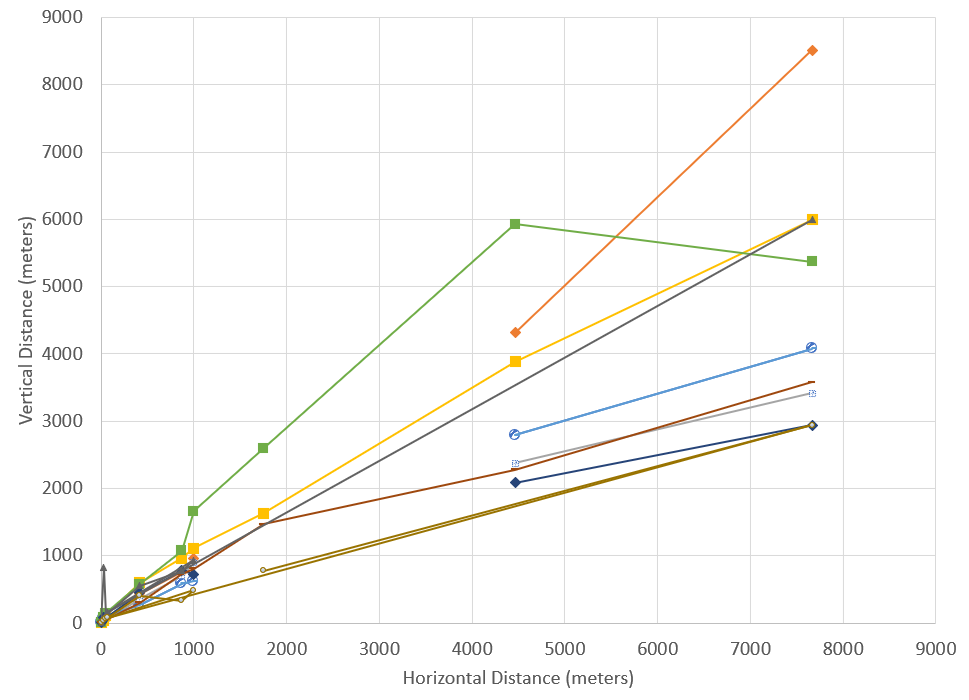}\\
    \caption{Ithaca Data: Note these physical horizontal distances and derived vertical distances} \label{IthFit}
    \vspace{2mm}
\end{figure}

For the data taken in Verplanck, the average celestial vault begins approximately 20~m overhead and distances to perceived different vertical points in the sky increases with increased horizontal distances. Similarly, results at Stewart Park show the celestial vault to begin approximately 10~m directly overhead. These data then show that as distance increases from the observer, the perceived height of the celestial vault also increases. These data are depicted in Figures~\ref{VerpFit} and~\ref{IthFit}. 
 
We began our experiment at Stewart Park in Ithaca, New York looking across the featureless vista of Cayuga Lake. Data were collected on clear, cloudless days. Observations were taken out to the most distant visible landmark, at a distance of approximately 8~km. However, preliminary review of these results was surprising because these data were not conforming to the expected model. 

Instead of appearing as the expected dome, the data showed the celestial vault to be lowest at the position directly overhead and to increase in height out to the most distant landmark. For that reason, we moved to a new location at Cortlandt Waterfront Park in Verplanck, New York where observations could be taken out to the horizon, approximately 13~km distance from the subject. At this location, more distant landmarks could be chosen.

\begin{table}
\begin{center}
 \begin{tabular}{ p{2.5cm}  p{5cm}  }
 
 \multicolumn{2}{c}{} \\
\hline
 Landmark Horizontal Distance (m) & Mean Vault Height (m) \\
 \hline
 \hline
 14.71 & 26 $\pm$ 10 \\
 17.31 & 32 $\pm$  11 \\
 66.1& 103 $\pm$  43 \\
 155 & 200 $\pm$ 120 \\
 249 & 320 $\pm$  120 \\
 1369 & 1400 $\pm$ 600 \\
 1725 & 1650 $\pm$ 820 \\
 4714 & 4300 $\pm$ 1900 \\
 10380 & 9200 $\pm$ 4000 \\
 12600 & 11000 $\pm$ 5100 \\
 \hline

\end{tabular}
\caption{Mean heights of the celestial vault as found in the Verplanck data.}\label{VerplankMean}
\end{center}
\end{table}

The trend is similar to that found in Ithaca.

\begin{table}
\begin{center}
 \begin{tabular}{ p{2.5cm}  p{5cm} }
 
 \multicolumn{2}{c}{}\\
\hline
 Landmark Horizontal Distance (m) & Mean Vault Height (m)\\
 \hline
 \hline
 3.26 & 7.3 $\pm$ 3.9 \\
 28.8 & 130 $\pm$  240 \\
 52.8 & 87 $\pm$  32\\
 418 & 420 $\pm$  120 \\
 868 & 730 $\pm$  200 \\
 999 & 880 $\pm$  330 \\
 4468 & 3300 $\pm$  1300 \\
 7667 & 4700 $\pm$  1800 \\
 \hline

\end{tabular}
\caption{Mean heights of the celestial vault as found in the Ithaca data.}\label{IthacaMean}
\end{center}
\end{table}


 \section{\label{sec:Analysis}Discussion}
 
 The resulting shapes of the celestial vault are in disagreement with \citet{Rees}, \citet{Kaufman}, and \citet{KG} in that they do not yield a spherical or flattened vault as expected.
 Given that the expectation is that the data should form the arc of a circle or an ellipse, we attempted to fit individual trials with the general equation of a conic section, but the small number of data points in each trial precluded any satisfactory such fits; all trials were just as consistent (in $r^2$) with a simple linear fit as they were with the quadratics. Therefore instead of attempting fits to individual trials, we opted to fit to a combined data set. 
 
\begin{figure}
    \centering
     \includegraphics[width=\linewidth]{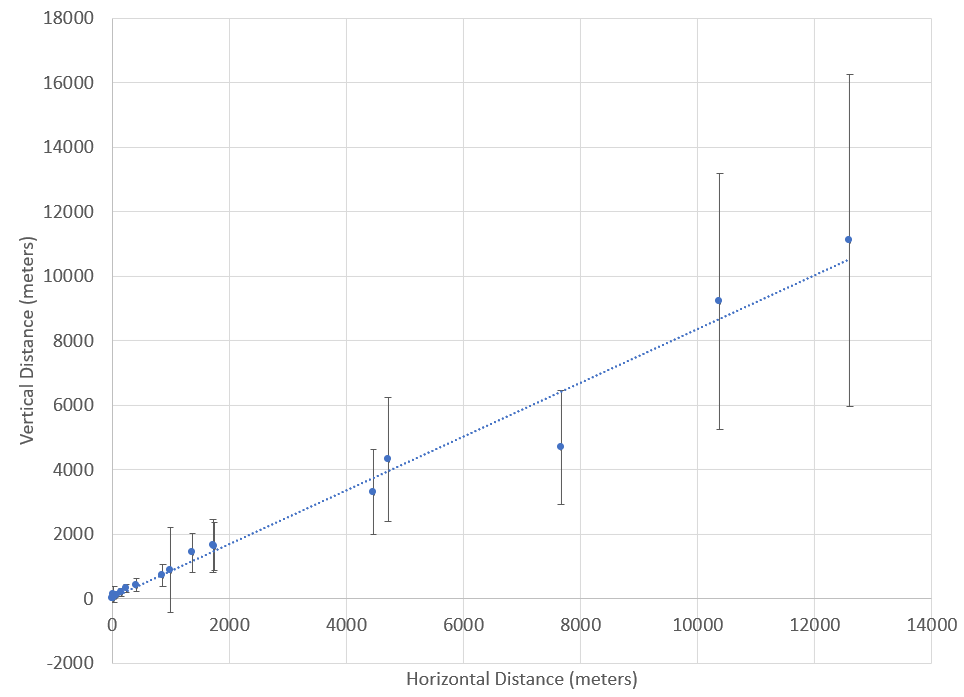}\\
    \caption{Linear fit to the complete data set, whereby horizontal distances are physical distances and vertical distances are derived distances.\label{linearAllData}}
    
\end{figure}

 This combined data set was constructed by finding, for each landmark, the mean and standard deviation of derived heights for all trials (See Tables~\ref{VerplankMean} and~\ref{IthacaMean}). We found a linear fit for this data set (See Figure~\ref{linearAllData}). The equation for this fit is: $y = 0.8339x + 35.663$~m, where $x$ is the horizontal distance in meters to each landmark and $y$ represents the average height of the celestial vault in meters. Interestingly, the $y$ intercept of this equation is 35.663 m, meaning that subjects on average perceive the zenith sky to be approximately 36 meters directly overhead. Compare to the typical distance to objects in the sky, such as clouds, which may be thousands of meters overhead in real distance. This low overhead altitude and steepness of the line makes it appear that subjects are not, for instance, measuring angles up to an imagined layer of clouds or any other horizontal plane as suggested by \citet{Rees}. 
 
\begin{figure}
    \centering
     \includegraphics[width=\linewidth]{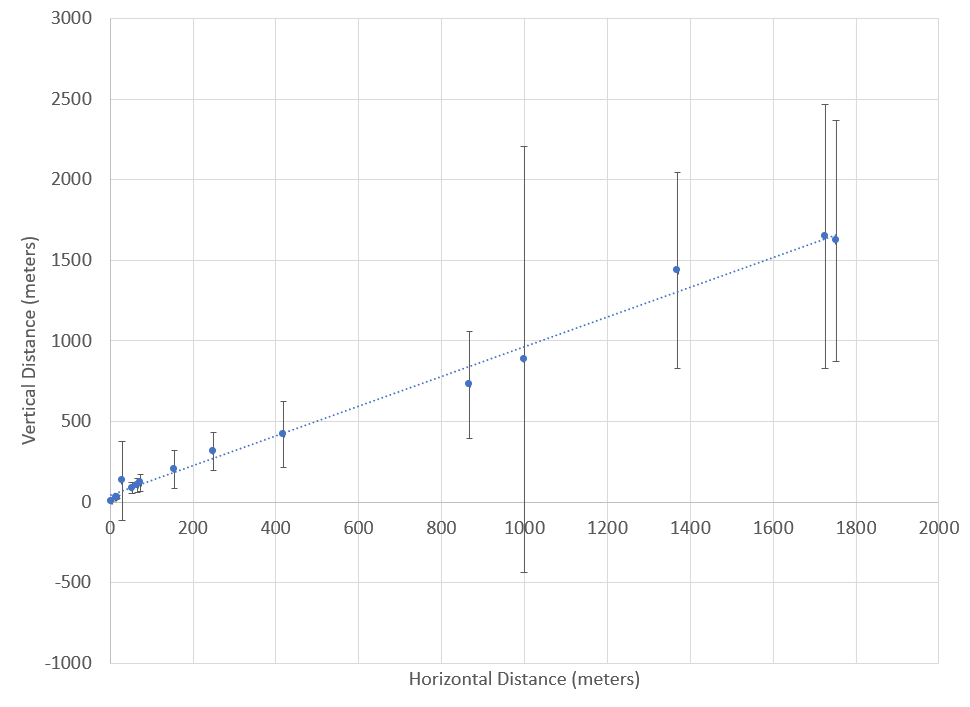}\\
    \caption{Linear fit to the data with landmark distances less than 2~km, whereby horizontal distances are physical distances and vertical distances are derived distances. \label{linearNearData}}
\end{figure}

\begin{figure}
    \centering
     \includegraphics[width=\linewidth]{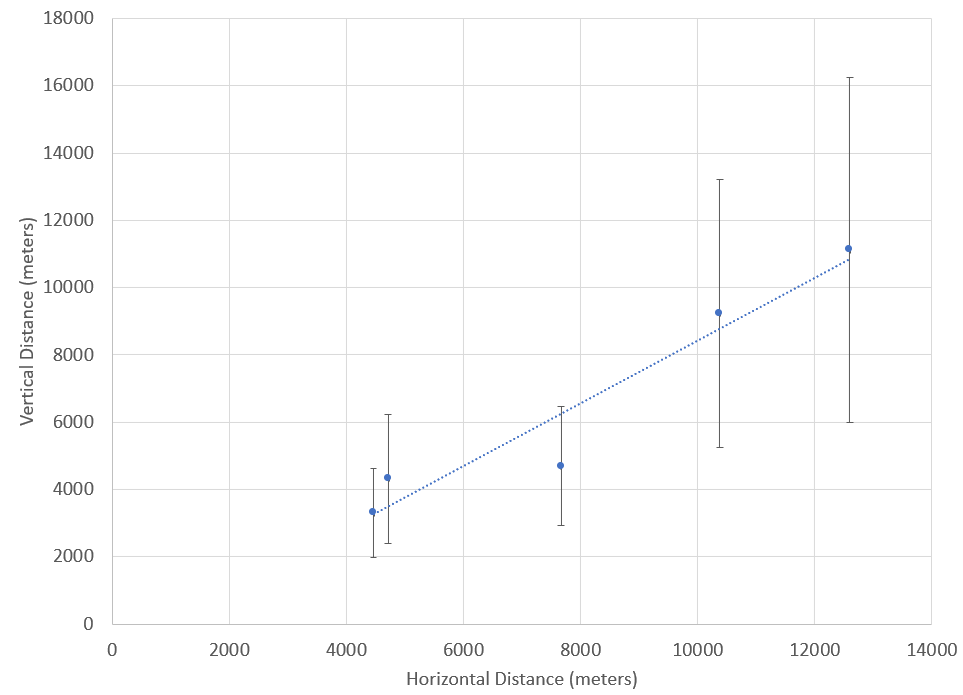}\\
    \caption{Linear fit to data with landmark distances greater than 2~km, whereby horizontal distances are physical distances and vertical distances are derived distances.\label{linearFarData}}
    
\end{figure}

 We also separately fit data for landmarks closer than, and farther than, 2~km (Figures~\ref{linearNearData} and~\ref{linearFarData}). The linear fit for the near data is: $y = 0.922x + 38.573$~m. Again, $x$ represents horizontal distance to each landmark in meters and $y$ represents the average perceived height of the sky in meters. The $y$ intercept of 38.573~m is similar to that of the total data fit intercept. The linear fit for the far data is: $y = 0.9321x + 893.21$~m. This tells us, from the slope and $y$ intercept of the far fit data, that celestial vault does not turn down to meet the horizon, but in fact does quite the contrary. At farther distances the vault surprisingly arcs more steeply upwards. 

 
Although we believe that each individual participant's responses generally agree with these averaged statements, the individual variation was large and the number of participants small. We also wanted to search for any consistency with the expected circular or elliptical shape of the arc of the vault. Thus we were interested in attempting to find a best \emph{shape} to which to fit the data. To explore these questions, we  constructed a separate scaled set of observations. Each observer’s data were scaled based on two separate attributes: the extreme points of the perceived height, and the slope of the linear fits of their perceived heights versus position. Each observer’s data were scaled such that each set would have the same slope of linear fit as all the others. We then attempted to fit the scaled data with general quadratic fits, in an effort to seek the best ellipse, circular arc, or parabola for the data sets.

Comparing the calculated height versus position data before and after scaling this way provides an indication of a potentially consistent shape of the celestial vault from person to person. While the data collected have uncertainties too large to provide an expression for the formula of this shape, it is suggestive that the best fit quadratic was generally an upward--facing parabola, and not an ellipse or ciruclar arc. Because of this similarity in the relationship recorded between subjects, and that no individual subject’s calculated height versus position graph ever touches the ground, it is possible to state that the data indicate that the celestial dome never contacts ground level. As a result, there is a region in the sky from the horizon to about 25 degrees of altitude which does not correspond to any point on our subjects’ mental model of the sky.

 \section{\label{sec:Conclusion}Conclusion}
Different experimental methods attempting to access the shape the perceived celestial vault arrive at different results. This may indicate that the hypothesis that a celestial vault exists at all is invalid.
One particularly puzzling aspect of our investigation is that, as measured using this technique, the celestial vault never curves downward to meet the horizon. If it had, as expected by the Ptolemaic flattened sky model, we should have expected subjects to point perfectly horizontally when asked to identify a position above an object on the horizon, but they did not. Indeed, when asked about the most distant landmarks, subjects typically indicated an altitude of between 20$^\circ$ and 45$^\circ$. The implication is that the perceived sky does not touch the horizon and there is an indeterminate space of 20$^\circ$ to 45$^\circ$ between perceived sky and ground.

20$^\circ$ is also the typical maximum altitude at which most Moon Illusion effects are claimed, and 20 to 45 degrees is the range of the bisecting angle found by Miller \& Neuberger, Angell, and others. In other words, the Moon Illusion primarily occurs in the indeterminate space between the ground and the observer's egocentric sky, i.e. the Moon is not then even upon the egocentric sky \emph{at all}. 
In such a position, there is no point on the ground associated with being directly underneath it. This may explain apocryphal notions of the Moon as ``hovering'' in front of the sky when the illusion is seen: the mind has no model on which to estimate its distance there. 

Our data show that on clear days, the perceived distance to the zenith is much closer to the observer than that to the horizon. Although significant differences exist among our many observers, almost all agree on this. This contradicts the Ptolemaic hypothesis of a flattened vault as described by Kaufman and Rock, but also the conclusions of Baird \& Wagner and To\v scovi\' c (who concluded that the vault is vertically elongated). Those studies both measured distances to objects as opposed to positions. Therefore, their methods were not capable of producing our outcome of an indeterminate gap between the ground and sky. Our results also indicate the celestial vault is only a few meters above the observer's head near the zenith. 

This experiment could be done with incrementally more accuracy by measuring angle of the pupil above the horizontal instead of the extended arm, but this would not affect the main qualitative result here, that there exists an indeterminate region of order $20^\circ$ altitude. This experiment could also be run in reverse, where participants are asked at night to point to a landmark below a given star. Once again, the perceived shape of the sky could be determined, compared to our current results and checked for consistency.

\section{Acknowledgements}

This work was supported by resources of the State University of New York at Cortland and Cornell University.



\begin{thebibliography}{}
\bibitem[Abhyankar(1989)]{Abhy}
Abhyankar, K. D.  BASI. 1989, 17, 55

\bibitem[Angell(1932)]{ANG}
Angell, Frank. ``Notes on the horizon illusion: II.'' The Journal of General Psychology 6.1, 1932: 133-156. 

\bibitem[Baird, Guliek, and Smith (1962)]{BGS62}
Baird, J. C., Gulick, W. L., and Smith, W. M. ``The effect of angle of regard on the size of after-images.'' Psychological Record, 1962: 12, 263-271. 

\bibitem[Baird and Wagner(1982)]{BW}
Baird, J. C., and Mark Wagner. ``The moon illusion: I. How high is the sky?'' Journal of Experimental Psychology: General 1982: 111.3:296. 

\bibitem[Desaguiliers (1736)]{Des}
Desaguiliers, J. T. RSPT, 1736, 39, 390-2. 

\bibitem[Goldstein(1962)]{Goldstein}
Goldstein, Gilbert. 1962, Sci, 138.3547, 1340-1341

\bibitem[Higashiyama and Adachi (2006)]{HAD}
Higashiyama, Atsuki, and Kohei Adachi. ``Perceived size and perceived distance of targets viewed from between the legs: Evidence for proprioceptive theory.'' Vision research 46.23 (2006): 3961-3976. 

\bibitem[Holway and Boring (1940a, 1940b)]{HB}
Holway, A. H., and Boring, E. G. (1940). ``The moon illusion and the angle of regard.'' The American Journal of Psychology, 53, 109–116. 

\bibitem[Jones and Wilson(2009)]{Jones}
Jones, Stephanie AH, and Alexander E. Wilson. ``The horizon line, linear perspective, interposition, and background brightness as determinants of the magnitude of the pictorial moon illusion.'' Perception and Psychophysics 71.1 (2009): 131-142. 

\bibitem[Kaufman and Kaufman(2000)]{Kaufman}
Kaufman, Lloyd, and James H. Kaufman. 2000, PNAS, 97.1, 500-505

\bibitem[Kaufman and Rock(1962)]{KAR}
Kaufman, Lloyd, and Irvin Rock. 1962, Sci, vol. 136, no. 3520, 953–61

\bibitem[Kim (2002)]{Kim}
Kim, N.-G. ``Oculomotor Effects in the Size-Distance Paradox and the Moon Illusion.'' Ecological Psychology 24, 2. 2012. 

\bibitem[King and Gruber(1962)]{KG}
King, William L., and Howard E. Gruber. 1962, Sci, 135.3509, 1125-1126

\bibitem[Liebowitz and Hartman(1959)]{Lieb}
Leibowitz, H., and T. Hartman. 1959, Sci, 130.3375, 569-570

\bibitem[Miller and Neuberger(1945)]{MN}
Miller, Albert, and Hans Neuberger. 1945, BAMS, 26.6, 212-216

\bibitem[Rees(1986)]{Rees}
Rees, W. G. 1986, QJRAS, 27, 205

\bibitem[Rock and Kaufman (1962)]{RK1} 
Rock, Irvin, and Lloyd Kaufman. 1962, Sci, vol. 136, no. 3521, 1023–31


\bibitem[Smith(1738)]{SM}
Smith, Robert. {\it A COMPLEAT SYSTEM OF OPTICKS: In Four BOOKS.
\/} Vol. 1. Cambridge: 1738. 


\bibitem[Suzuki (2007)]{Suzu}
Suzuki, Kotaro. ``The moon illusion: Kaufman and Rock's (1962) apparent‐distance theory reconsidered 1.'' Japanese Psychological Research 49.1 (2007): 57-67. 

\bibitem[Taylor and Boring(1942)]{TB}
Taylor, Donald W., and Edwin G. Boring. ``The moon illusion as a function of binocular regard.'' The American Journal of Psychology 55.2 (1942): 189-201. 

\bibitem[To\v skovi\' c (2009)]{T1}
To\v skovi\' c (2009). 2009, POBeo, 86, 385-389

\bibitem[Wallach, Boring, Rock, and Kaufman(1962)]{Wallach}
Wallach, H., Boring, E. G., Rock, I., and Kaufman, L. 1962, Sci, 137.3533, 902-906

\bibitem[Wolbarsht and Lockhead(1985)]{WL}
Wolbarsht, Myron L., and Gregory R. Lockhead. 1985, ApOpt, 24.12, 1844-1847

\bibitem[Zehender(1900)]{LH}  
Zehender, W. von (1900). ``Die Form des Himmelsgewolbes und <las
Grosser­Erscheinen der Gestirne am Horizont.'' Zeitschrift fur
Psychologie, 24, 218-84. 

\end{thebibliography}
\end{document}